\documentclass[aps,prx,preprint,letterpaper,groupedaddress,longbibliography]{revtex4-2}
\usepackage{amsmath,amsthm,amsfonts}
\theoremstyle{definition}

\usepackage{multirow}
\usepackage{array}
\usepackage{soul}
\usepackage{float}
\usepackage{graphicx}
\usepackage{afterpage}

\usepackage[font=small,labelsep=quad,format=plain]{caption}
\usepackage{subcaption}
\usepackage[colorlinks=true,citecolor=blue,linkcolor=blue,urlcolor=blue]{hyperref}
\usepackage{cleveref}
\crefname{equation}{Eq.}{Eqs.}
\crefname{section}{Sec.}{Secs.}
\crefname{subsection}{Sec.}{Secs.}
\crefname{appendix}{Appendix}{Appendices}
\crefname{figure}{Fig.}{Figs.}
\crefname{table}{Table}{Tables}
\crefname{proposition}{}{}
\crefname{corollary}{}{}

\usepackage{bm}
\usepackage{physics}
\usepackage[dvipsnames]{xcolor} 

\newcommand{\eunits}{kV cm\textsuperscript{-1}}
\newcommand{\mobunits}{cm\textsuperscript{2} V\textsuperscript{-1} s\textsuperscript{-1}}
\newcommand{\diffunits}{cm\textsuperscript{2} s\textsuperscript{-1}}
\newcommand{\qnumber}{\lambda}
\newcommand{\qnumberp}{\lambda'}
\newcommand{\qnumbertext}{\lambda}

\begin{document}

\title{High-field charge transport and noise in \textit{p}-Si from first principles}

\author{David S. Catherall}
\affiliation{Division of Engineering and Applied Science, California Institute of Technology, Pasadena, CA 91125, USA}
\author{Austin J. Minnich}
\email{aminnich@caltech.edu}
\affiliation{Division of Engineering and Applied Science, California Institute of Technology, Pasadena, CA 91125, USA}
\date{\today}

\begin{abstract}
    The parameter-free computation of charge transport properties of semiconductors is now routine owing to advances in the ab-initio description of the electron-phonon interaction. Many studies focus on the low-field regime in which the carrier temperature equals the lattice temperature and the current power spectral density (PSD) is proportional to the mobility. The calculation of high-field transport and noise properties offers a stricter test of the theory as these relations no longer hold, yet few such calculations have been reported. Here, we compute the high-field mobility and PSD of hot holes in silicon from first principles at temperatures of 77 and 300 K and electric fields up to 20 \eunits~along various crystallographic axes. We find that the calculations quantitatively reproduce experimental trends including the anisotropy and electric-field dependence of hole mobility and PSD. The experimentally observed rapid variation of energy relaxation time with electric field at cryogenic temperatures is also correctly predicted. However, as in low-field studies, absolute quantitative agreement is in general lacking, a discrepancy that has been attributed to inaccuracies in the calculated valence band structure. Our work highlights the use of high-field transport and noise properties as a rigorous test of the theory of electron-phonon interactions in semiconductors.
\end{abstract}

\maketitle
\newpage

\section{Introduction}
The calculation of semiconductor charge transport properties from first principles is of fundamental interest as a test of the theory of electron-phonon interactions and of practical interest for device applications \cite{Ponce:2020,Bernardi:2016}. Recent advances in the ab-initio description of electron-phonon interactions have enabled the calculation of phonon-limited mobility without adjustable parameters using density functional theory (DFT), density functional perturbation theory (DFPT), Wannier interpolation to fine grids needed for transport calculations, and the Boltzmann transport equation (BTE) \cite{Giustino:2017,Bernardi:2016}. Implementations of this approach are now available in various software packages \cite{Smidstrup:2019,EPW,ABINIT,Zhou:2021,Protik:2022}, and have been applied to calculate low-field properties in a number of materials such as  Si \cite{Li:2015,Fiorentini:2016,Ponce:2021} and GaAs \cite{Ponce:2021,Zhou:2016,Liu:2017}, 2D materials including graphene \cite{Park:2014,Borysenko:2010,Gunst:2016,Restrepo:2014} and MoS\textsubscript{2} \cite{Restrepo:2014,Li:2015,Gunst:2016,Kaasbjerg:2012,Sohier:2018}, and others \cite{Ponce:2020}. The approach continues to develop, with advances in the ab-initio description of two-phonon scattering \cite{Lee:2020}, neutral and ionized impurity scattering \cite{Lu:2020, Lu:2022}, quadrupole interactions \cite{Park:2020,Brunin:2020}, and others \cite{Desai:2021,Jhalani:2020}.\par

In addition to low-field transport, calculations of high-field transport and noise properties are of particular interest as additional phenomena occur that are not present in the low-field regime. For instance, at sufficiently large electric fields the conductivity exhibits a field dependence and an anisotropy even in cubic crystals \cite{Schmidt:1963}. Furthermore, the fluctuation-dissipation theorem is not valid in non-equilibrium systems, meaning that fluctuation and noise properties such as the current power spectral density contain qualitatively distinct information about transport processes not available from mean properties such as mobility \cite{Nougier:1980,Hartnagel:2001,Lax:1960,Reggiani:1985}. In non-polar semiconductors, these high-field effects were first experimentally observed as a departure from Ohm's law in Ge and Si \cite{Ryder:1951,Ryder:1953,Shockley:1951,Zucker:1961}. Development of the time-of-flight (TOF) technique allowed the hot-carrier drift velocity to be measured, revealing high-field effects including anisotropy, drift velocity saturation, and negative differential conductivity in \textit{n}-Si \cite{Canali:1975} and \textit{p}-Si \cite{Norris:1967,Canali:1974,Ottaviani:1975,Reggiani:1986}. The TOF technique has also been used to measure the diffusion coefficient \cite{Sigmon:1969,Jacoboni:1979,Nava:1979,Brunetti:1981}. The contemporaneous development of  measurements of the spectral noise power of current fluctuations also allowed for the determination of noise temperature \cite{Erlbach:1962,Nougier:1973} and  diffusion coefficient \cite{Nava:1979,Bareikis:1980,Brunetti:1981,Gasquet:1986} using the fluctuation-diffusion relation ~\cite{Gantsevich:1979}. Measurements at cryogenic temperatures have also found that non-Ohmic behaviour becomes more apparent as temperature decreases \cite{Asche:1970,Canali:1971,Tschulena:1972}.



Various theoretical and numerical methods have been employed to interpret these measurements in terms of microscopic transport processes \cite{Conwell:1967,Jacoboni:1983, Nougier:1994,Reggiani:1997_2,Jacoboni:1979}. In particular, high-field transport phenomena in \textit{p}-Si have been primarily studied via Monte Carlo (MC) methods, which have investigated both steady-state \cite{Fischetti:1991,Dewey:1993,Rodriguez:2005,Soares:2020} and fluctuation phenomena \cite{Nava:1979,Reggiani:1986,Reggiani:1997_1}. While these and other works employed semi-empirical approximations for the band structure and scattering rates, recent works  have employed full-band calculations which partially relax prior approximations \cite{Hess:1991,Yoder:1994_2,Shichijo:1981,Jungemann:1999, Fitzer:2003} and applied them to various materials including \textit{p}-Si  \cite{Bude:1994,Fischetti:1996,Jallepalli:1997,Bufler:1998_1,Fischer:2000,Nguyen:2003}. However, no MC study on \textit{p}-Si has employed a fully ab-initio band structure and scattering rates.



The fully ab-initio treatment of electron-phonon scattering \cite{Bernardi:2016,Li:2021} has recently been used to calculate low-field hole properties in silicon \cite{Ma:2018,Ponce:2018}. These works found that ab-initio calculations with one-phonon scattering, spin-orbit coupling, and the relaxation time approximation is generally adequate to predict the low-field mobility. Overestimates of the mobility of around 30\% were attributed to inaccuracies in the DFT valence band structure. Recent works have extended the ab-initio method to study high-field transport phenomena and noise \cite{Choi:2021,Cheng:2022,Maliyov:2021}. High-field transport calculations offer a stricter test of the theory because band anisotropy, intervalley and interband scattering, and energy relaxation take on increased importance \cite{Reggiani:1980,Conwell:1967,Hartnagel:2001,Jacoboni:1980}. Additionally, the full solution to the BTE is necessary at high fields \cite{Lundstrom:2000,Choi:2021}. However, recent methods for the ab-initio treatment of high-field transport \cite{Cheng:2022, Choi:2021, Maliyov:2021} have not yet been applied to \textit{p}-type semiconductors.\par

Here, we report ab-initio calculations of high-field mobility and current PSD of hot holes in \textit{p}-Si at electric fields up to 20 and 12 \eunits~and temperatures of 300 and 77 K. We find that the calculated properties quantitatively reproduce the trends of high-field transport including the electric field-dependence and anisotropy of the mobility and PSD, and the rapid variation of energy relaxation time with electric field at cryogenic temperatures. We find that the absolute transport properties are uniformly overestimated by around $\sim 25$\%, consistent with the origin being inaccuracies in the valence band structure. Our work demonstrates the utility of the first-principles calculations of high-field transport and fluctuation properties of semiconductors as a stringent test of the theory of electron-phonon interactions.

\section{Theory and numerical methods}

The details of the method used in this work have been described previously \cite{Choi:2021, Cheng:2022}. In summary, high-field transport and noise properties are calculated by solving the Boltzmann transport equation with the electronic states, phonon dispersion, and electron-phonon collision matrix computed from first principles. In a homogeneous system in which carriers are subjected to an electric field, the BTE is given by:

\begin{equation}
    {
        \frac{q\mathbf{E}}{\hbar}
        \cdot
        \nabla_\mathbf{k}
        f_{\qnumber}
        =
        -\Theta_{\qnumber\qnumberp}
        \Delta f_{\qnumberp}
    }
    \label{shortBTE}
\end{equation} 

\noindent where $q$ is the carrier charge, $\mathbf{E}$ is the electric field vector, and $f_{\qnumber}$ is the carrier occupation function indexed by $\qnumbertext$ which represents the combined indices of band $n$ and wave vector $\mathbf{k}$. Here, $\Theta_{\qnumber\qnumberp}$ is the linearized collision integral as given by Eq. 3 of Ref.~\cite{Choi:2021}. The collision integral depends on the phonon populations which may be perturbed by Joule heating, an effect known as the hot phonon effect \cite{Kocevar:1980,Paranjape:1968}. Owing to the small free carrier densities in the relevant experiments ($\lesssim10^{14}\,\mathrm{cm}^{-3}$), this effect may be neglected as the non-equilibrium phonon generation rate is too small to affect hole transport properties.

Equation \ref{shortBTE} can be expressed as a linear system of equations as described in Sec. II.A. of Ref.~\cite{Choi:2021}. The only necessary modification for the present work is a change of sign for the term implementing the momentum-space derivative to account for the charge carriers being holes. The BTE then takes the form:

\begin{equation}
    {
        \sum_{\qnumberp}
        \Lambda_{\qnumber\qnumberp}
        \Delta f_{\qnumberp}
        =\sum_\gamma
        \frac{e\mathrm{E}_\gamma}{k_BT}
        v_{\qnumber,\gamma}
        f^0_\qnumber
    }
    \label{fullBTE}
\end{equation}

\noindent where $\mathrm{E}_\gamma$ and $v_{\qnumber,\gamma}$ are the electric field strength and hole velocity in the $\gamma$ Cartesian axis, and $\Delta f_{\qnumber}$ is the perturbation to the equilibrium distribution function $f^0_{\qnumber}$. The relaxation operator $\Lambda_{\qnumber\qnumberp}$ is defined as:

\begin{equation}
    {
        \Lambda_{\qnumber\qnumberp}
        =
        \Theta_{\qnumber\qnumberp}
        -\sum_\gamma
        \frac{e\mathrm{E}_\gamma}{\hbar}
        D_{\qnumber\qnumberp,\gamma}
    }
    \label{relop}
\end{equation}

\noindent where $D_{\qnumber\qnumberp,\gamma}$ is the finite difference matrix representation of the momentum-space derivative. The solution to the linear system, $\Delta f_{\qnumber}$, can be used to calculate various observables. For instance, the mobility is given as \cite{Li:2015}:

\begin{equation}
    {
        \mu_{\alpha\beta}(\mathbf{E})
        =
        \frac{2e^2}{k_BT\mathcal{V}}
        \sum_{\qnumber}
        v_{\qnumber,\alpha}
        \sum_{\qnumberp}
        \Lambda^{-1}_{\qnumber\qnumberp}
        \left(
            v_{\qnumberp,\beta}
            f^0_{\qnumberp}
        \right)
    }
    \label{mobility}
\end{equation} 

\noindent where $\mathcal{V}$ is the supercell volume, $\alpha$ is the direction along which the current is measured, and $\beta$ is the direction along which the electric field is applied. In addition to mobility, which represents a mean characteristic of the steady-state distribution, the current power spectral density (PSD), which quantifies the random fluctuations of carriers about the non-equilibrium steady-state distribution, may be calculated using the BTE. As given in Sec.~II.B. of Ref.~\cite{Choi:2021}, the PSD can be computed as:

\begin{equation}
    {
        S_{j_\alpha j_\beta}(\mathbf{E},\omega)
        =
        2\left(
            \frac{2e}{\mathcal{V}}
        \right)^2
        \mathfrak{R}
        \left[
            \sum_{\qnumber}
            v_{\qnumber,\alpha}
            \sum_{\qnumberp}
            \left(
                i\omega\mathbb{I}
                +\Lambda
            \right)_{\qnumber\qnumberp}^{-1}
            \left(
                f^s_{\qnumberp}
                \left(
                    v_{\qnumberp,\beta}
                    -V_\beta
                \right)
            \right)
        \right]
    }
    \label{vPSD}
\end{equation}

\noindent where $\omega$ is the angular frequency, $\mathbb{I}$ is the identity matrix, and  $f^s_{\qnumber}=f^0_{\qnumber}+\Delta f_{\qnumber}$ is the steady distribution. Here, $V_\beta$ is the drift velocity, given by:

\begin{equation}
    {
        V_\beta
        =
        \frac{1}{N}
        \sum_{\qnumber}
        v_{\qnumber,\beta}
        f^s_{\qnumber}
    }
    \label{aveV}
\end{equation}

\noindent where $N=\sum_{\qnumber}f_{\qnumber}$ is the number of holes in the Brillouin zone.


The numerical methods are identical to those in Refs.~\cite{Choi:2021, Cheng:2022}. The calculation of electronic structure and electron-phonon matrix elements are performed with Quantum Espresso \cite{Giannozzi:2009,Giannozzi:2017} using a coarse 8$\times$8$\times$8 grid, wave function energy cutoff of 60 Ryd, and a lattice constant of 5.430 \AA. The spin-orbit interaction is included. The interpolation to fine grids is performed using PERTURBO \cite{Zhou:2021}. The BTE, formulated as a linear system of equations, is solved using the GMRES algorithm \cite{Scipy}. Observables are calculated using a Brillouin zone sum. Transport and noise properties are calculated for lattice temperatures of 300 K and 77 K, with carrier densities of 10\textsuperscript{16} cm\textsuperscript{-3} and 10\textsuperscript{14} cm\textsuperscript{-3}, maximum energy of 192 meV and 92 meV, Gaussian smearing parameters of 5 meV and 2.5 meV, and grid densities of 120\textsuperscript{3} and 180\textsuperscript{3}, respectively. Increasing the grid densities to 130\textsuperscript{3} and 190\textsuperscript{3}, respectively, resulted in less than a 1\% change in mobility and PSD with the exception of the 77 K PSD, which exhibited a 3\% change at the highest fields. These convergence results are in agreement with a recent low-field study \cite{Ma:2018}. Only the heavy and light valence bands were included in the calculations presented as the split-off band was found to contribute negligibly to all quantities while considerably increasing computational costs. In the following results, normalized computed quantities are shown relative to their value at 1 V cm\textsuperscript{-1}. The error between calculations and experiments for arbitrary property $y$ as a function of some parameter is defined as $||y_{exp}-y_{calc}||_2\,/\,||y_{exp}||_2 $. The anisotropy for a property along different crystallographic directions is defined at one parameter value $x$ for direction $\alpha$ as $|{y(x)}_{[100]}-{y(x)}_{\alpha}|\,/\,|{y(x)}_{[100]}|$.

\section{Results}

\subsection{Transport}

We begin by examining the electric field dependence of the DC mobility at 300 K. The calculated low-field mobility is 525 \mobunits, overestimated compared to the experimental value of $\sim 450$ \mobunits~\cite{Nava:1979,Canali:1971}. We find that calculations yield the same overestimate at all fields, so the mobilities have been normalized to facilitate the comparison of trends; computed data are normalized to the calculated low-field mobility, while experimental data are normalized to the low-field mobility measured by Ref.~\cite{Nava:1979} ($450$ \mobunits).\par

\begin{figure}[t]
    \centering{
        \phantomsubcaption\label{fig:300K_mob}
        \phantomsubcaption\label{fig:300K_KDE}
        \includegraphics[width=6.5in, height=3.65in]{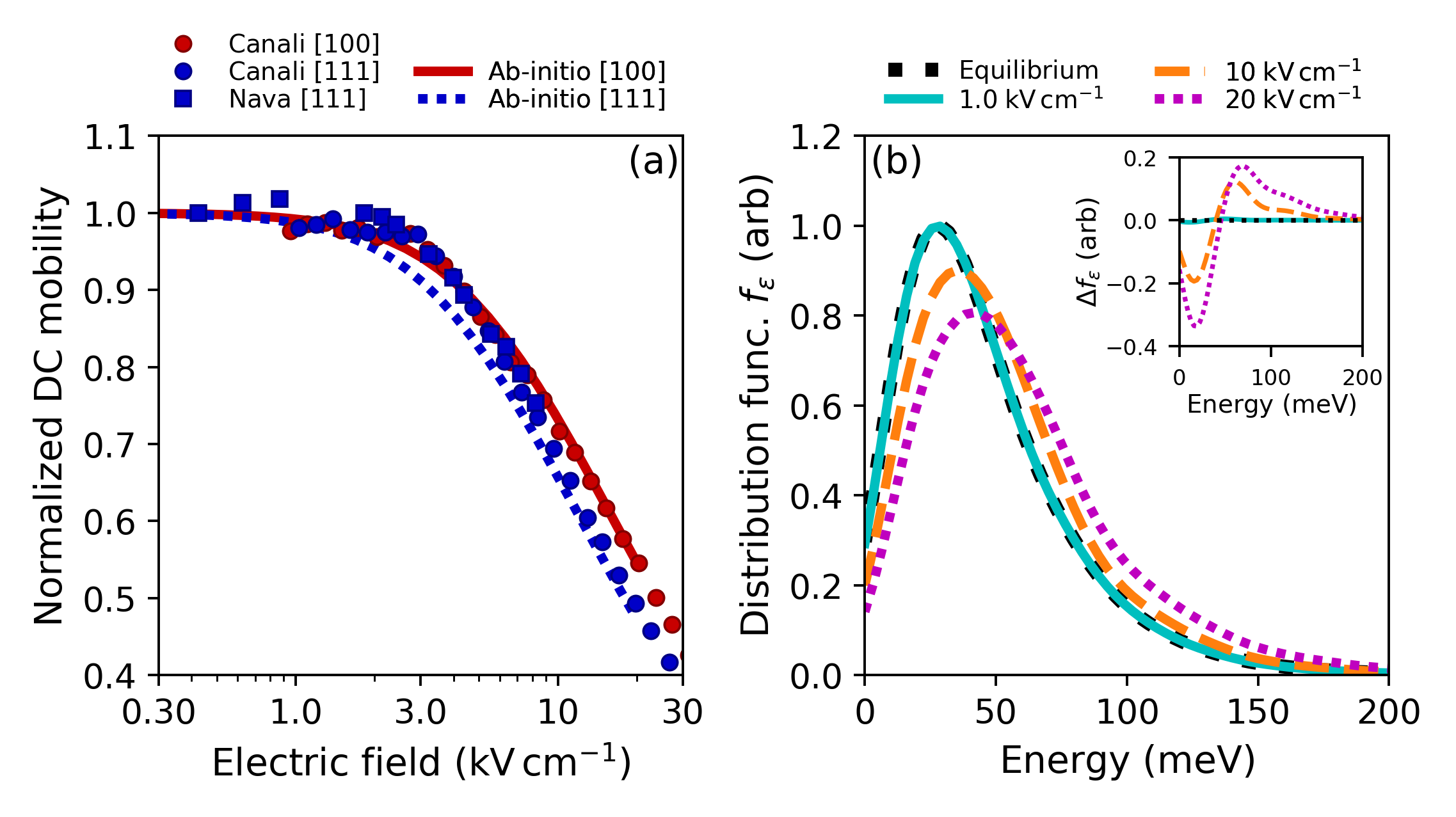}}
    \caption{(a) Normalized DC mobility  versus electric field at 300 K. Calculated data normalized to their values at 1 V cm\textsuperscript{-1}, experimental data normalized to the low field value of Ref.~\cite{Nava:1979} ($450$ \mobunits). Calculated data shown for electric fields in various crystallographic axes including $[100]$ (solid red line) and $[111]$ (dotted blue line). Experimental data shown from Canali et al.  \cite{Canali:1971} ($[100]$ as red circles, $[111]$ as blue circles), and Nava et al.  \cite{Nava:1979} ($[111]$ as blue squares). (b) Hole distribution function versus energy ($f_\epsilon$), from the valence band maximum at 300 K for electric fields of 1 \eunits (solid light blue line), 10 \eunits~ (dashed orange line), and 20 \eunits~(dotted magenta line) applied along the $[100]$ crystallographic axis. The equilibrium distribution is shown as a dashed black line. The distribution function was calculated using kernel density estimation. Inset: deviation from the equilibrium distribution function $\Delta f_\epsilon=f_\epsilon-f^0_\epsilon$. The redistribution of holes to higher energies with increased field strength is evident, leading to the decrease in mobility in (a).}

    \label{fig:300K}
\end{figure}

The calculated DC mobility versus electric field in various crystallographic axes is shown in \cref{fig:300K_mob} along with experimental data from Refs.~\cite{Canali:1971,Nava:1979}. At low fields $\lesssim 1$ \eunits, the mobility is nearly constant as expected in the Ohmic regime. At fields $\gtrsim 1$ \eunits, the mobility decreases with increasing field. We find the normalized mobilities are in quantitative agreement with experiment, with errors of only 2.5\% and 5.4\% for the $[100]$ and $[111]$ directions, respectively (compared to Ref. \cite{Canali:1971}). Additional data can be found in Ref.~\cite{Ottaviani:1975}, which is not shown but is also in similar agreement. There is a minor discrepancy in the onset of anisotropy, which occurs at fields $\sim 80\%$ lower than predicted by experiment for the $[111]$ direction. However, the anisotropy at high fields is in good agreement with experiment; at 20 \eunits~the $[111]$ anisotropy is 13\% and 11\% for the calculations and measurements, respectively. It was also found that the properties along the $[110]$ and $[111]$ directions are degenerate at all fields, in agreement with data from Ref.~\cite{Ottaviani:1975}.

The decrease in mobility with increasing field can be be attributed to the heating of the holes owing to their finite energy relaxation time. Hole scattering rates increase with energy (see Fig.~6b of Ref.~\cite{Ma:2018}). Thus, the mobility decreases as holes are heated by the applied electric field. To illustrate this point, we computed the hole distribution versus energy at various electric fields in \cref{fig:300K_KDE}. At low fields ($\lesssim 3$ \eunits), the steady-state distribution of holes coincides with the thermal distribution, and $\Delta f_\epsilon\approx0$. At higher fields, $\Delta f_\epsilon$ becomes non-negligible and the steady-state distribution $f_\epsilon$ is shifted to higher energies, ultimately leading to a nonlinear response between drift velocity and electric field.

\begin{figure}[t]
    \centering{
        \phantomsubcaption\label{fig:77K_mob_a}
        \phantomsubcaption\label{fig:77K_mob_b}
        \includegraphics[width=6.5in, height=3.25in]{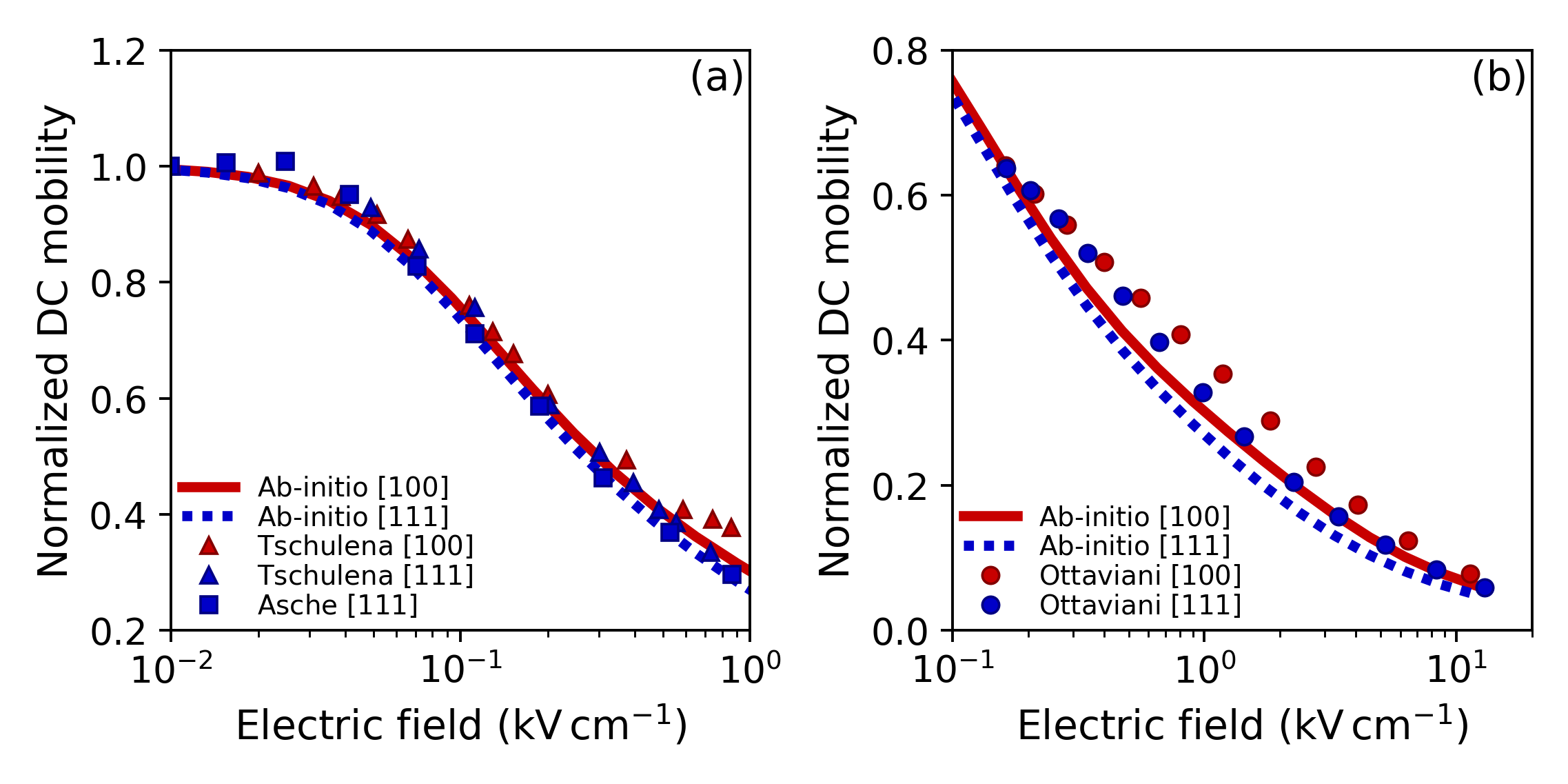}}
    \caption{(a) Normalized DC mobility  versus electric field at 77 K. Calculated data shown for electric fields along the crystallographic axes $[100]$ (solid red line) and $[111]$ (dotted blue line). Calculated data normalized to their values at 1 V cm\textsuperscript{-1}. Experimental data shown from Tschulena  \cite{Tschulena:1972} ($[100]$ red triangles, $[111]$ blue triangles) and Asche et al.  \cite{Asche:1970} ($[111]$ blue squares). (b) Same as (a) but for data from Ref.~\cite{Ottaviani:1975} ($[100]$ red circles, $[111]$ blue circles), normalized to 9800 \mobunits.}
    \label{fig:77K_mob}
\end{figure}

Next, we show the mobility versus electric field at 77 K in \cref{fig:77K_mob}. Data from Refs.~\cite{Tschulena:1972,Asche:1970} were originally reported in normalized form, and are thus presented separately from Ref.~\cite{Ottaviani:1975} for which the absolute high-field mobility was reported but not the low-field mobility, complicating the normalization. At this temperature the low-field mobility is again overestimated, with a value of 12300 \mobunits~compared to the experimental value of $\sim 9800$ \mobunits~\cite{Ottaviani:1975,Logan:1960}. Mobility versus field at 77 K for Refs.~\cite{Tschulena:1972,Asche:1970} are presented in \cref{fig:77K_mob_a}, along with calculations. Quantitative agreement with experiment is observed from 0 -- 0.6 \eunits. Above this field there is a minor discrepancy in the $[100]$ direction from Ref.~\cite{Tschulena:1972}, but other data remain in agreement with calculations. The overall error between calculations in the $[100]$ direction with data from Ref. \cite{Tschulena:1972} is 3.6\%.\par

Experimental data from Ref.~\cite{Ottaviani:1975}, which do not reach the Ohmic regime, have been normalized to 9800 \mobunits~and are presented in \cref{fig:77K_mob_b} along with calculated data. While the computed $[111]$ anisotropy is once again in good quantitative agreement, being 20\% (15\%) for calculations (measurements) at 12 \eunits, the mobility-field values only agree qualitatively, with an error of $\sim$12.5\% in all crystallographic directions. At 77 K, the $[110]$ mobility is not degenerate with the $[111]$ mobility, with 23 \% (29\%) calculated (measured) anisotropy, but data for this field direction is omitted for clarity. It should be noted, however, that comparing experimental data from both figures in the region of 0.1--0.7 \eunits, the data from Refs.~\cite{Tschulena:1972,Asche:1970} agree quantitatively with the calculations while the data from Ref.~\cite{Ottaviani:1975} do not. Some inconsistency therefore exists between measurements in these references, and the appropriate reference for comparison is not presently clear.

\subsection{Noise}

We next examine the PSD in the low-frequency limit, defined as frequencies for which $\omega \tau \ll 1$. Here, $\tau$ is a characteristic relaxation time of energy, momentum, or similar parameter. In this limit, the current PSD is proportional to the diffusion coefficient through the fluctuation-diffusion relation (see Eqn.~2.49 of Ref.~\cite{Gantsevich:1979}). Experimentally, the diffusion coefficient has been obtained by the TOF method \cite{Reggiani:1980} and from low-frequency noise measurements \cite{Nava:1979,Gasquet:1986}. We calculate the diffusion coefficient through the current PSD, which was computed at 1 GHz to obtain the low-frequency value; negligible differences were observed between 1 GHz and 100 MHz.\par

\begin{figure}[t]
    \centering{
        \phantomsubcaption\label{fig:diffusion_a}
        \phantomsubcaption\label{fig:diffusion_b}
        \includegraphics[width=6.5in, height=3.25in]{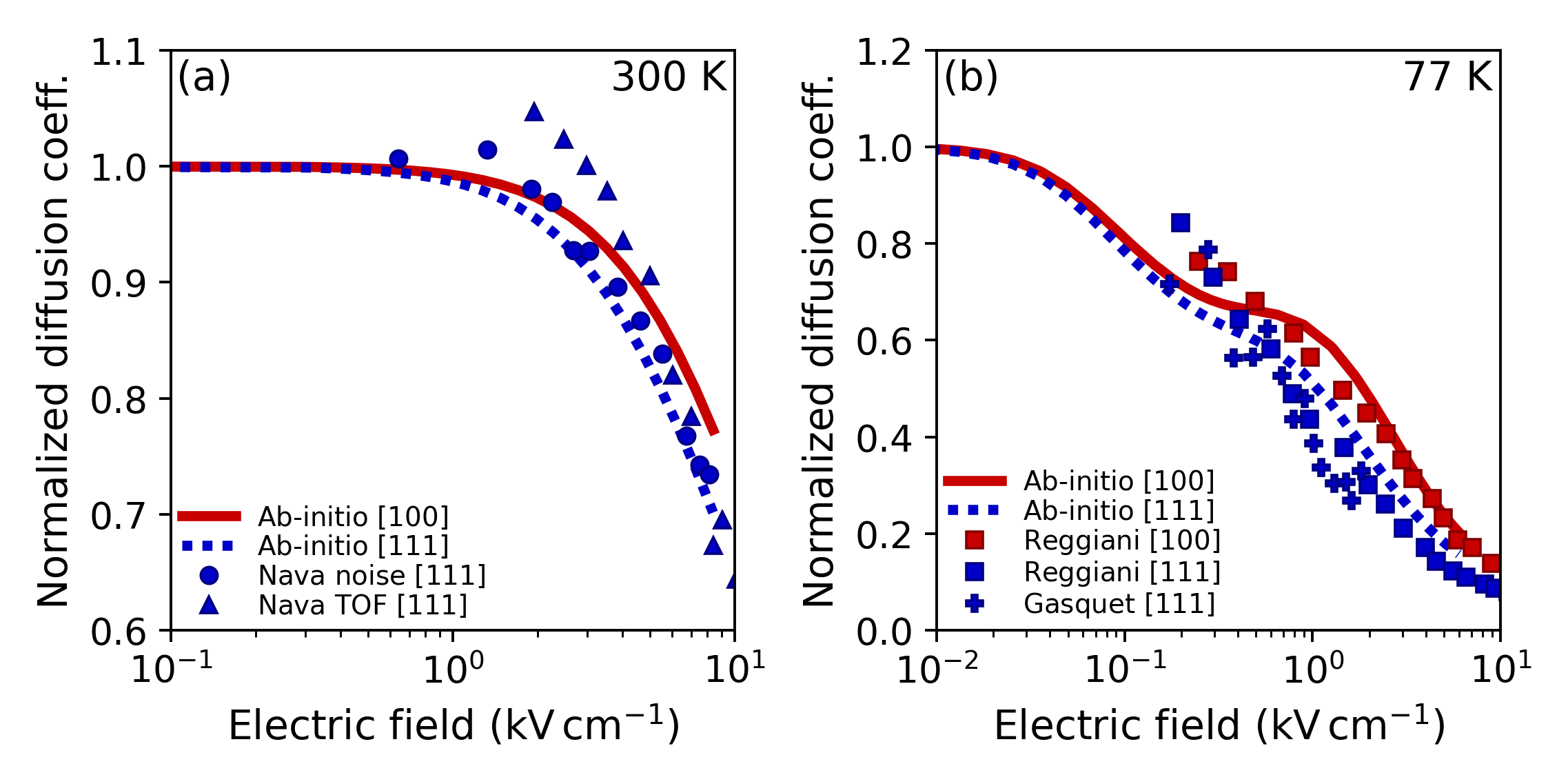}}
    \caption{(a) Normalized hole diffusion coefficient versus electric field at 300 K. Calculated data normalized to their values at 1 V cm\textsuperscript{-1}. Calculated data shown for electric fields applied in the crystallographic axes $[100]$ (solid red line) and $[111]$ (dotted blue line). Experimental data shown from Nava et al. \cite{Nava:1979} ($[111]$, noise measurements as blue circles, TOF measurements as blue triangles) and normalized to 11.6 \diffunits. (b) Same as (a) at 77 K. Experimental data shown from Reggiani et al. \cite{Reggiani:1980_2} ($[100]$ as red squares, $[111]$ as blue squares) and Gasquet et al. \cite{Gasquet:1986} ($[111]$ as blue crosses).}
    \label{fig:diffusion}

\end{figure}

The hole diffusion coefficient versus electric field at 300 K is shown in \cref{fig:diffusion_a}. Experimental measurements from TOF and noise data from Ref.~\cite{Nava:1979} are also shown normalized to the equilibrium value reported by the same study (11.6 \diffunits). The calculations are in reasonable quantitative agreement with experiment, with an error of 5.5\%, although the data from noise measurements are much closer to calculations, with an error of only 2.5\%. At 300 K, comparing \cref{fig:300K_mob,fig:diffusion_a}, we observe that the decrease in diffusion coefficient with increasing electric field follows the same trend as that of the mobility. This result is expected as the system remains in the warm carrier regime for all fields considered. Specifically, at the highest field presented here, the mean energy of the steady-state distribution is  greater than that of the equilibrium distribution by $\sim 10\%$, indicating that the steady-state distribution remains close to the equilibrium distribution. In this case, the fluctuation-dissipation relation remains approximately valid \cite{Choi:2021}, so $D(\mathbf{E})\approx\mu(\mathbf{E})k_BT/q$ according to the Einstein relation.\par

At 77 K and high fields ($\gtrsim 0.1$ \eunits), holes enter the hot carrier regime, with mean energies exceeding the equilibrium value by up to 200\% at 6 \eunits. Thus, in contrast to the 300 K case, it is expected that the mobility and diffusion coefficient exhibit distinct dependencies on electric field. We present the diffusion coefficient versus field at 77 K in \cref{fig:diffusion_b}. Experimental data have been normalized to the approximate low-field diffusion coefficient obtained from the Einstein relationship assuming 9800 \mobunits~as the low-field mobility (yielding 65 \diffunits). The calculations predict an initial decrease, followed by a plateau around $\sim 0.5$ \eunits, and a subsequent decrease. This trend has been previously attributed to the rapid increase in energy relaxation rate once optical phonon emission becomes possible (see Ref. \cite{Zakhleniuk:1989} and p.70 of Ref.~\cite{Hartnagel:2001}). The differing trends of mobility and diffusion coefficient highlight the failure of the Einstein relationship and hence the utility of calculating both steady-state and fluctuation properties for high fields.\par

The 77 K diffusion coefficient calculations exhibit qualitative agreement with experiment, but discrepancies are clearly present. At 3 \eunits, the calculated (experimental) $[111]$ anisotropy is 26\% (39\%), indicating the anisotropy has been reasonably captured. However, the plateau feature in the calculation is not evident in the data. We note that experimental details are not available for data from Ref. \cite{Reggiani:1980_2}, and data from both Refs. \cite{Reggiani:1980_2,Gasquet:1986}, obtained from TOF measurements, have uncertainties on the order of 25\% \cite{Canali:1975_2}. Comparison of the trends is therefore challenging considering the uncertainties in experiment.

\subsection{Energy relaxation time}

At low temperatures the mechanisms by which carriers lose energy depend on the electric field strength. At low fields, scattering is dominated by acoustic phonons because holes lack sufficient energy to emit optical phonons, and the energy relaxation time (ERT) achieves its maximum value with little dependence on field strength. At high fields, the carriers are heated to energies sufficient to emit optical phonons, and the ERT decreases by orders of magnitude \cite{Dargys:1972}. Experimentally, the field-dependent ERT has been estimated by measuring the transverse noise temperature and using a phenomenological energy balance equation as:

\begin{equation}
    \tau_\epsilon(\mathrm{E})
    =
    \frac{3 k_B}{2 e}
    \frac{\left(T^n_\perp-T_0 \right)}{\mu \mathrm{E}^2}
    \label{etau_aveE}
\end{equation}

\noindent where $\mu$ is the DC mobility at electric field $E$, $T^n_\perp$ is the transverse noise temperature, and $T_0$ is the lattice temperature (see  Sec.~9.2 of \cite{Hartnagel:2001} or Sec.~4.5 of Ref.~\cite{Conwell:1967}). However, computationally the energy relaxation time can be attained more straightforwardly though the energy PSD; this property follows the form of the current PSD in \cref{vPSD} but with the quantity of interest being energy instead of velocity, as follows:

\begin{equation}
    S_{\epsilon\epsilon}(\mathbf{E},\omega)
    =
    2\left(
        \frac{2}{\mathcal{V}}
    \right)^2
    \mathfrak{R}
    \left[
        \sum_{\qnumber}
        \epsilon_{\qnumber}
        \sum_{\qnumberp}\left(
            (i\omega\mathbb{I}+\Lambda)^{-1}_{\qnumber,\qnumberp}
            (f^s_{\qnumberp}(\epsilon_{\qnumberp}-\mathcal{E}))
        \right)
    \right]
\end{equation}

\noindent where $\epsilon_{\qnumber}$ is the energy of a hole relative to the energy at the valence band maximum, and $\mathcal{E}$ is the mean energy given by:

\begin{equation}
    {
        \mathcal{E}
        =
        \frac{1}{N}
        \sum_{\qnumber}
        \epsilon_{\qnumber}
        f^s_{\qnumber}
    }
    \label{aveE}
\end{equation}

\noindent A representative calculated energy PSD is shown in Fig. 3b of Ref.~\cite{Choi:2021}. The ERT is then obtained from a Lorentzian fit of the computed energy PSD \cite{Hartnagel:2001}:

\begin{equation}
    S_{\epsilon\epsilon}(\mathbf{E},\omega)
    =
    \frac{S_{\epsilon\epsilon}(\mathbf{E},0)}{1+\left(\omega\tau_\epsilon\right)^2}
    \label{etau_EPSD}
\end{equation}

\begin{figure}[t]
    \centering{
        \includegraphics[width=3.375in, height=3.25in]{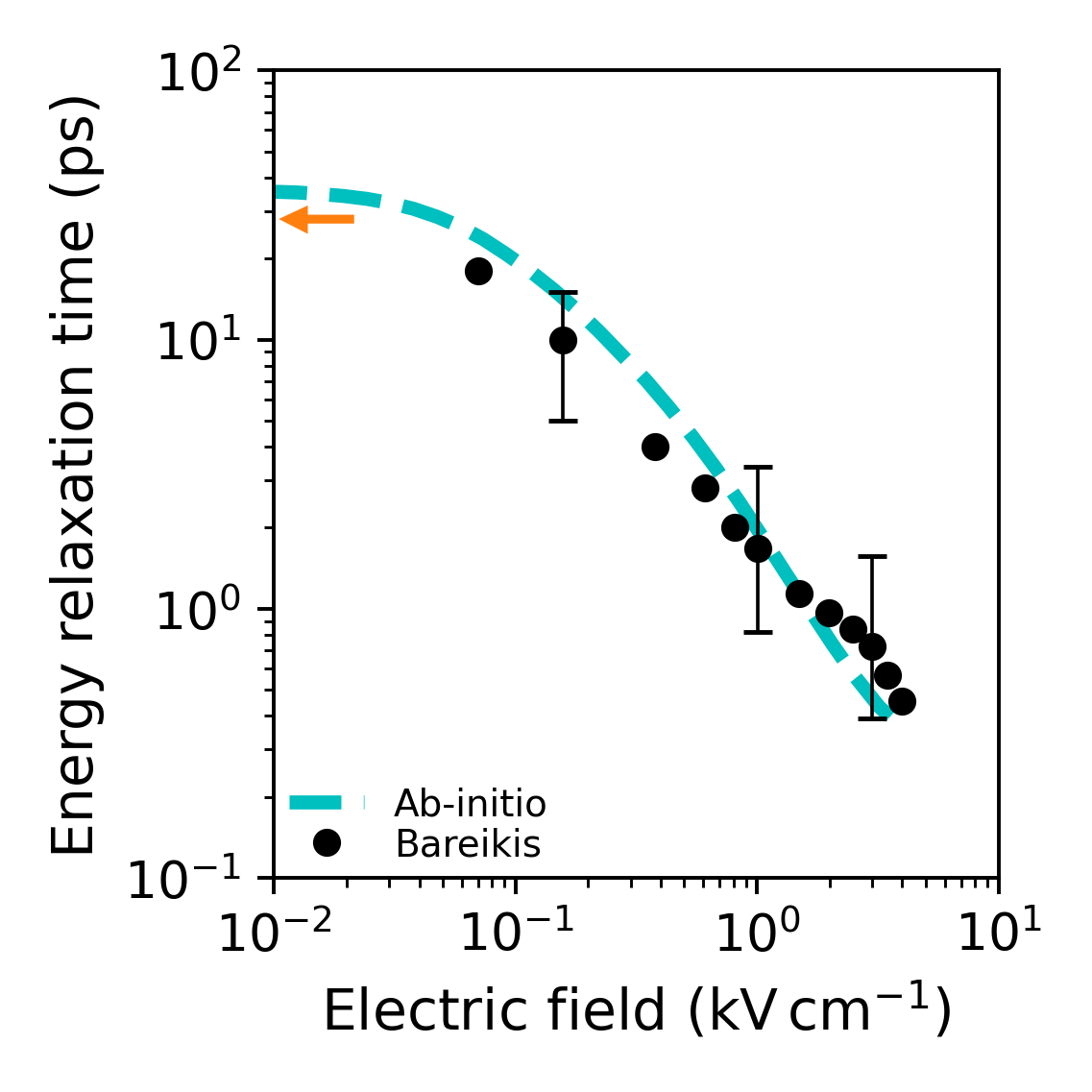}}
    \caption{Hole energy relaxation time versus electric field applied in the $[110]$ direction at 77 K from calculation (dashed light blue line) and experiment (black circles, Ref.~\cite{Bareikis:1980}). Also shown is the equilibrium (zero field) value of $\tau_\epsilon$ measured by Hess and Seeger \cite{Hess:1969} (orange arrow).}
    \label{fig:77K_ERT}
\end{figure}

We compare our ERT calculations to the experimental data reported in Ref.~\cite{Bareikis:1980}. The experimental and calculated ERT are shown in \cref{fig:77K_ERT}. We find the equilibrium energy relaxation time at zero electric field to be within $\sim 30$\% of the value obtained from \cite{Hess:1969}. High-field calculations are also in quantitative agreement with experiment considering the experimental uncertainty. As expected, the ERT shows a rapid decrease with field once optical phonon emission becomes possible around $\sim 0.1$ \eunits, consistent with the start of the diffusion coefficient plateau in \cref{fig:diffusion_b}.

\section{Discussion}

The quantitative agreement of most normalized transport properties with experiment indicates that momentum and energy relaxation scattering processes of hot holes in Si are adequately described by the lowest level of pertubation theory in the electron-phonon interactions used in prior works for non-polar semiconductors \cite{Li:2015,Ponce:2018,Ma:2018}. Additionally, the relative differences in curvature of the valence band structure in different crystallographic directions appear to be captured as evidenced by the agreement of the high-field anisotropy. However, as in low-field studies \cite{Ponce:2018,Ma:2018}, absolute transport properties are uniformly overestimated by $\sim 25$\% at all fields. The low-field mobility overestimate has previously been attributed to an inaccurate valence band structure \cite{Ponce:2018}. Examination of our valence band structure and that of Ref.~\cite{Ponce:2018} (Fig. 2b, `SOC' case) indicates that the two are in quantitative agreement to within 5\% as measured by the average root-mean square difference of the dispersions, excepting the heavy holes in the $\Gamma- \mathrm{K}$ direction which exhibit a $\sim  20$\% heavier mass compared to those in Ref.~\cite{Ponce:2018}. This difference is several times smaller than the difference with the dispersion which yields the experimental mobility. Our computed scattering rates are in similar quantitative agreement with prior literature, with the values falling within the range of those reported in Fig.~6b of Ref.~\cite{Ma:2018}. Therefore, our results support the findings from low-field studies in which the overestimated transport properties were attributed to inaccuracies in the DFT valence band structure.




Future work may provide additional tests of the theory by using the present formalism to compute other transport properties, including the high-field drift velocity at $T<40$ K for which a plateau in drift velocity versus electric field is observed;  \cite{Ottaviani:1975} and the microwave PSD, which exhibits a non-monotonic feature at 77 K \cite{Bareikis:1980}. A superlinear current-voltage characteristic has also been reported at for \textit{p}-Si at 77 K under uniaxial stress \cite{Ross:1970}, which has been attributed to the shifting of carriers to the light holes due to strain-induced splitting of the heavy and light-hole bands at $\mathbf{k}=0$. This phenomenon makes an investigation of the transport and noise properties of strained \textit{p}-Si of distinct interest.

\section{Summary}

We have presented a study of hot hole transport and noise in silicon. We find that the ab-initio calculations quantitatively reproduce various experimental trends up to 20 \eunits. Absolute properties are generally overestimated by $\sim 25$\%, consistent with prior low-field studies. This agreement may be improved by the use of a more accurate valence band structure. This work highlights the use of high-field transport and noise properties as a rigorous test of the theory of electron-phonon interactions in semiconductors.

\section{Acknowledgements}

This work was supported by the National Science Foundation under Award No. 1911926. The authors thank A. Choi, B. Hatanp\"a\"a, P. Cheng, S-N. Sun, and J. Sun for code development and discussions.

\bibliography{bib}

\end{document}